\documentclass[fleqn,usenatbib, nofootnoteinbib]{mnras}
\usepackage[utf8]{inputenc}  
\usepackage[T1]{fontenc}
\usepackage{graphicx,psfrag}
\usepackage{mathrsfs}
\usepackage{amsmath,amsfonts,amssymb,pifont,gensymb}
\usepackage{multirow,enumerate}
\usepackage{comment,hyperref}
\usepackage{color}
\usepackage{acronym}
\usepackage{xspace}
\usepackage[normalem]{ulem}
\usepackage{mathtools}
\usepackage{subfigure}
\usepackage{makecell}
\usepackage{newtxtext,newtxmath}
\usepackage{changepage}
\usepackage{appendix}
\usepackage{lipsum}

\usepackage{geometry}
\newacro{BH}{black hole}
\newacro{NS}{neutron star}
\newacro{PN}{Post-Newtonian}
\newacro{BBH}{binary black hole}
\newacro{BNS}{binary neutron star}
\newacro{EOB}{effective-one-body}
\newacro{NR}{numerical relativity}
\newacro{GW}{gravitational wave}
\newacro{EOS}{equation-of-state}

\newcommand{\be}{\begin{equation}}
\newcommand{\ee}{\end{equation}}
\newcommand{\bea}{\begin{eqnarray}}
\newcommand{\eea}{\end{eqnarray}}
\newcommand{\bel}{\begin{align}}
\newcommand{\eel}{\end{align}}

\newcommand{\dett}{\text{det}}

\newcommand{\boldtheta}{\boldsymbol{\theta}}
\newcommand{\boldLambda}{\mathbf{\Lambda}}

\def\i{{\rm i}}

\def\GMc2{{\rm G M_{\odot} c^{-2}}}

\def\SEOBNRv4T{\texttt{SEOBNRv4T}\xspace}

\usepackage{color}
\definecolor{cyan}{rgb}{0,0.9,0.9}
\definecolor{orange}{rgb}{0.9,0.5,0}
\definecolor{magenta}{rgb}{1,0,1}
\definecolor{purple}{rgb}{0.8,0.4,0.8}
\definecolor{gray}{rgb}{0.5,0.5,0.5}
\definecolor{mygreen}{rgb}{0.1,0.8,0.1}
\definecolor{darkblue}{rgb}{0.0,0.0,0.6}

\title{A fast and precise methodology to search for and analyse strongly lensed gravitational-wave events} %

\author[J. Janquart et al.]{
Justin Janquart,$^{1, 2}$\thanks{E-mail: j.janquart@uu.nl}
Otto A. Hannuksela,$^{1, 2}$
Haris K.$^{1, 2}$
and  Chris Van Den Broeck$^{1, 2}$
\\
$^{1}$Nikhef – National Institute for Subatomic Physics, Science Park, 1098 XG Amsterdam, The Netherlands\\
$^{2}$Institute for Gravitational and Subatomic Physics (GRASP), Department of Physics, Utrecht University, Princetonplein 1, 3584 CC Utrecht, The Netherlands
}

\date{\today}

\pubyear{2021}

\begin{document}

\maketitle

\begin{abstract}
\noindent
Gravitational waves, like light, can be gravitationally lensed by massive astrophysical objects such as galaxies and galaxy clusters.
Strong gravitational-wave lensing, forecasted at a reasonable rate in ground-based gravitational-wave detectors such as Advanced LIGO, Advanced Virgo, and KAGRA, produces multiple images separated in time by minutes to months. These images appear as repeated events in the detectors: gravitational-wave pairs, triplets, or quadruplets with identical frequency evolution originating from the same sky location. To search for these images, we need to, in principle, analyse all viable combinations of individual events present in the gravitational-wave catalogues. An increasingly pressing problem is that the number of candidate pairs that we need to analyse grows rapidly with the increasing number of single-event detections. At design sensitivity, one may have as many as $\mathcal O(10^5)$ event pairs to consider. To meet the ever-increasing computational requirements, we develop a fast and precise Bayesian methodology to analyse strongly lensed event pairs, enabling future searches. 
The methodology works by replacing the prior used in the analysis of one strongly lensed gravitational-wave image by the posterior of another image; the computation is then further sped up by a pre-computed lookup table. 
We demonstrate how the methodology can be applied to any number of lensed images, enabling fast studies of strongly lensed quadruplets.
\end{abstract}


\section{Introduction}
\label{sec:Intro}
Gravitational waves (GWs) result from cataclysmic events that distort the fabric of space and time. 
The Advanced LIGO \citep{TheLIGOScientific:2014jea} and Advanced Virgo \citep{TheVirgo:2014hva} detectors have found dozens of GW signals emitted by the mergers of binary neutron stars or black holes \citep{LIGOScientific:2018mvr,Abbott:2020niy}. Advanced LIGO and Virgo are scheduled to be upgraded further. Meanwhile, Japan's KAGRA~\citep{Somiya:2011np, Aso:2013eba,Akutsu:2018axf,Akutsu:2020his} has become online. 
As the sensitivities of these instruments improve, many novel avenues in research could become observationally feasible. One such avenue could be the study of GW lensing.

%%%-------------------------------------------------------------------------
When GWs travel from their source to the Earth, they can be gravitationally lensed by intervening objects (galaxies, galaxy clusters, and other compact objects)~\citep{Ohanian1974, Degushi1986, Wang:1996as, Nakamura1998, Takahashi:2003ix, Dai:2017huk, Ezquiaga:2020gdt, Oguri:2018muv, Liu:2020par, Lo:2021nae}. Lensing could produce several observable effects on GW signals detectable by ground-based GW detectors. A small fraction of binary black hole mergers will be strongly lensed by intervening galaxies~\citep{Dai:2016igl, Ng:2017yiu, Li:2018prc, Oguri:2018muv} or galaxy clusters~\citep{Smith:2017mqu,Smith:2018gle,Smith:2019dis,Robertson:2020mfh,Ryczanowski:2020mlt}, producing multiple images detectable as repeated events~\citep{Wang:1996as,Haris:2018vmn}. The images can be magnified by the lens, resulting in a biased luminosity distance measurement~\citep{Dai:2016igl,Ng:2017yiu,Pang:2020qow}, or inverted along one or both principal axes, resulting in an overall shift in the GW phase~\citep{Dai:2017huk,Ezquiaga:2020gdt}, and they arrive a few minutes up to years apart. However, because the GW wavelength is typically much smaller than the lens size, lensing does not affect the signal's frequency content, referred to as the geometrical optics limit~\citep{Takahashi:2003ix}. Thus, GWs strongly lensed by galaxies or galaxy clusters appear as repeated events with identical frequency evolution and sky location but separated in time. If smaller compact objects (stars, black holes) intervene with the GW, microlensing can also occur. In that case, the geometrical optics approximation can break down, and frequency-dependent ``beating patterns'' to the waveform appear~\citep{Takahashi:2003ix,Cao:2014oaa,Lai:2018rto,Christian:2018vsi,Singh:2018csp,Hannuksela:2019kle,Meena:2019ate,Pagano:2020rwj,Cheung:2020okf,Kim:2020xkm}. 

%%%------------------------------------------------------------------------------
If identified, lensed GWs could give rise to several exciting possibilities.  For example, they might enable us to locate merging black holes at a sub-arcsecond precision by matching the GW image properties with the properties of the lenses discovered in electromagnetic surveys when quadruple images are available~\citep{Hannuksela:2020xor}. Similar localisation might be possible with galaxy cluster lensing even when only a signal double is found, owing to the rarity of the cluster lenses~\citep{Smith:2018gle,Ryczanowski:2020mlt} (see also Refs.~\citep{Sereno:2011ty, Yu:2020agu}). When accompanied by an electromagnetic counterpart, they could enable precision cosmography studies owing to the sub-millisecond lensing time-delay measurements granted by GW observations~\citep{Sereno:2011ty, Liao:2017ioi, Cao:2019kgn, Li:2019rns}. Additionally, lensed GWs might enable improved GW propagation tests by comparing the lensing time delay of strongly lensed waves with their transient electromagnetic counterparts to measure the speed of gravity relative to the speed of  light~\citep{Baker:2016reh, Fan:2016swi}. Moreover, strongly lensed events allow us to detect the same event multiple times at different detector orientations, effectively multiplying the number of detectors by the number of images to arrive at an enlarged (synthetic) detector network; this could be exploited to probe the full GW polarisation content, including alternative polarizations~\citep{Goyal:2020bkm}. Another prospective avenue is detecting intermediate-mass and primordial black holes through microlensing observations~\citep{Lai:2018rto,Jung:2017flg,Diego:2019rzc,Oguri:2020ldf}. 

%%%--------------------------------------------------------------------------
The principal idea to search for strong lensing is to locate event pairs (and from there, possibly triplets and quadruplets) with similar detector-frame parameters arriving at the detector from the same sky location. To assess if an event is strongly lensed, we compare the likelihood that an event pair is lensed against the likelihood that they were produced by astrophysical coincidence. Two parameter-estimation-based approaches exist to do this. The first one is the posterior-overlap methodology, which performs a Gaussian kernel density estimation (KDE)-based fit on the single event posterior density functions. It then tests if any given event pair is consistent with lensing by assessing the consistency of the posteriors~\citep{Haris:2018vmn}. This approach allows for rapid tests on large quantities of data. The second approach, joint parameter estimation analysis, abandons the KDE-based fits in favour of sampling the full joint likelihood, enabling improved accuracy~\citep{Liu:2020par, Lo:2021nae}. 

%%%--------------------------------------------------------------------------
An increasingly pressing problem is the rising computational demand of these strong lensing analyses. As an order-of-magnitude estimate, at design sensitivity, the number of observed GW events is likely to reach $\mathcal{O}(10^3)$~\citep{Ng:2017yiu},which results in $\sim 5\times 10^{5}$ event pairs. Even after dropping event pairs which show significant mismatch in the inferred parameters, there would be a large number of event pairs that need to be analysed.
To address the rising computational demand, we develop a new method: A fast and precise way to analyse strongly lensed GW event pairs by using the posterior of one event of the event pairs as a prior for the analysis of the other event in an importance sampling procedure.  
We further accelerate the methodology with a \emph{fast likelihood computation}. The methodology fills the gap between posterior-overlap~\citep{Haris:2018vmn} and joint parameter estimation~\citep{Liu:2020par,Lo:2021nae} methodologies in terms of speed and precision, enabling fast but still relatively accurate strong lensing analyses. 

%%%--------------------------------------------------------------------------
We summarise the relevant strong lensing theory and how it manifests itself in the detectors in Sec.~\ref{sec:GWlensing}. We describe the Bayesian framework and the approximations involved in our methodology to study strongly lensed events in Secs.~\ref{sec:StrongLensHypo},~\ref{sec:ConditionedEvidence}, and~\ref{sec:BayesianFrame}. We provide an example of a typical use case by analysing a non-lensed and a lensed pair in Sec.~\ref{sec:TypicalAnalysis}. We detail how one can use our approach to analyse multiple images in Sec.~\ref{sec:MultipleImages} and demonstrate an example application in Sec.~\ref{sec:FourImageExample}. A summary, conclusions, future outlook are given in Sec.~\ref{sec:Conclusions}. 

%%%--------------------------------------------------------------------------
\section{Strongly lensed gravitational waves}
\label{sec:GWlensing}
Strong lensing splits the GWs into multiple potentially observables images, which are categorised by their image type. Type-I images correspond to the minimum of the Fermat potential and leave the overall shape of the GWs unaffected. Type-II images are saddle points of the potential and Hilbert-transform the GWs. Finally, Type-III images correspond to the maximum of the Fermat potential and invert the original image and thus the waveform. Such images are typically captured by an overall phase shift of the waveform, referred to as the Morse phase~\citep{Dai:2017huk, Ezquiaga:2020gdt}. 
Besides the potential image inversions and Hilbert transforms, the images can also be magnified. 

%%%--------------------------------------------------------------------------
The lens is often a galaxy or galaxy cluster. Galaxy lensing typically produces two or four Type-I/II bright images (although a fifth, Type-III bright central image can be observed in rare cases when the density slope of the galaxy is very shallow~\citep{Collett:2015tma,2013ApJ...773..146D, Collett:2017ksf}) separated by a few minutes to months~\citep{Ng:2017yiu, Haris:2018vmn}.  

%%%--------------------------------------------------------------------------
Compared to galaxy lenses, galaxy cluster lenses can have much more complex lens morphologies, and therefore can produce a much richer spectrum of images, separated by up to years~\citep{Smith:2017mqu,Smith:2018gle,Smith:2019dis,Robertson:2020mfh,Ryczanowski:2020mlt}. 
Irrespective of the lens configuration, each lensed image will arrive at the detector as a Type-I, Type-II, or Type-III image, separated by times $t_j$, and have their intensities magnified by factors $\mu_j$. 
Thus, the search approach is usually phenomenological and agnostic to the specific lens configuration that produced the images. 

%%%--------------------------------------------------------------------------
However, we note that the lensed time delays and image types can be forecasted statistically, at least for galaxy lenses~\citep{Oguri:2018muv, Haris:2018vmn}.  
Such statistical information would improve the discriminatory power of our searches, and we expect to pursue this in future work. 

%%%---------------------------------------------------------------
\begin{figure}
    \centering
    \includegraphics[keepaspectratio, width=0.5\textwidth]{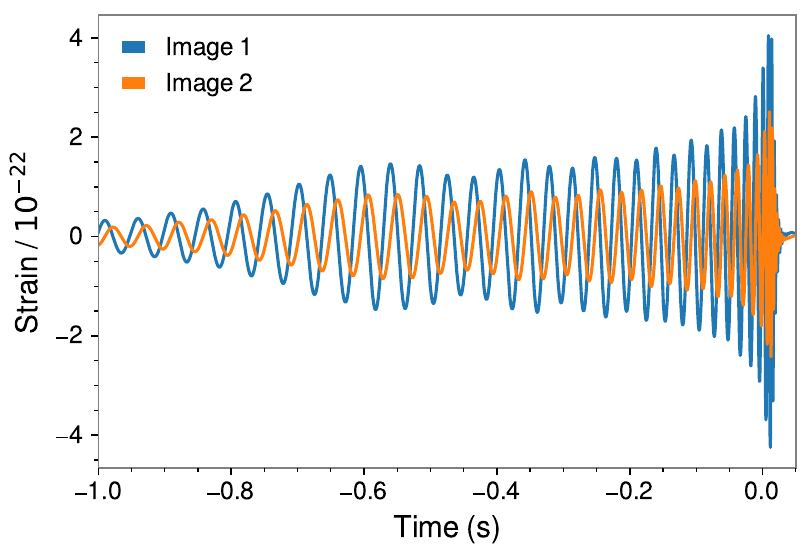}
    \caption{Two strongly lensed gravitational-wave images of a single binary black hole event. 
      The lens magnifies the two waves differently, and that is why their amplitudes are different. 
      One of the images has also been inverted by a lens, and thus there is a relative phase difference between the waveforms. 
      Normally, the two images would arrive at different times. 
      However, here the two waveforms have been time-shifted to superpose them.}
    \label{fig:lensedWF}
\end{figure}

%%%---------------------------------------------------------------------
Consequently, in the frequency domain, the GW waveform of the $j^{th}$ image is modified such that it arrives with a time delay $t_j$, experiences an overall magnification $\mu_j$ due to the focusing by lensing,\footnote{The magnification factor corresponds to the inverse of the determinant of the lensing Jacobian matrix~\citep{Schneider:1992, Haris:2018vmn}.} and can be inverted/Hilbert-transformed, an effect captured in an overall complex ``Morse factor'' (phase shift) $n_j$~\citep{Dai:2017huk,Ezquiaga:2020gdt}:
\begin{equation}\label{eq:lensed_waveform}
   \tilde{h}_{L}^j(f; \boldtheta, \mu_j, t_{j}, n_j) = \sqrt{\mu_{j}} e^{(2\pi i f  t_{j} - i\pi n_{j} \text{sign}(f))}  \tilde{h}_{U}(f; \boldtheta) \, ,
\end{equation}
where
$\tilde{h}_{U}(f;\boldtheta)$ is the frequency-domain waveform in the absence of lensing with the usual set of binary parameters $\boldtheta$, and $\tilde{h}_{L}^j(f; \boldtheta, \mu_j,  t_{j}, n_j)$ is the lensed frequency-domain waveform; for an illustration of a lensed waveform, see Fig.~\ref{fig:lensedWF}.

Note that the image parameters enter the GW waveform in an overall multiplicative factor, which can easily be decoupled from the rest of the waveform. 
For brevity, we will denote the individual image properties (magnifications, time delays, Morse factors) by $\boldLambda_j$.
An overview of the notations used in this work can be found in Table~\ref{tab:notations}.

\begin{table*}
    \centering
    \begin{tabular}{l l l}
        \hline
        \hline
        \textbf{Notation} &  \textbf{Set of parameters} & \textbf{Description} \\
        \hline
        \hline
         $\boldtheta$ & $\{m_{1}, m_{2}, a_{1}, a_{2}, d_{L}, t_{c}, \dots\}$ & Orbital BBH parameters  \\
         $\mathbf{\Theta}$ & $\{ m_{1}, m_{2}, a_{1}, a_{2}, d^{\mathrm{obs},1}_{L}, t_{c}^{\mathrm{obs},1}, n_1, \dots \}$ & The observed BBH parameters and the Morse phase $n_1$ of the \emph{reference image} (image 1) \\
         $\boldLambda_{j}$ & $\{\mu_{j}, t_{j},  n_{j}\}$ & Lensed image properties \\
         $\mathbf{\Phi}_j$ & $\{ \mu_{\rm rel}^{j, 1}, t_{j1}, n_{j1}\}$& Relative image properties (note that we define $\mathbf{\Phi}=\mathbf{\Phi}_1$ for convenience) \\
         \hline
    \end{tabular}
    \caption{Overview of the different vectors used in this work and their corresponding set of parameters and descriptions.}
    \label{tab:notations}
\end{table*}

%%%--------------------------------------------------------------------------
We can absorb the $\mu_j^{1/2}$ into an observed luminosity distance $d_L^{\rm obs, j}$ and the time-delay due to lensing into an observed coalescence time $t_c^{\rm obs, j}$, such that
\begin{align} \label{eq:effective_params}
    d_L^{\rm obs, j} &= \frac{d_L}{\sqrt{\mu_j} }\,,\\
    t_c^{\rm obs, j} &= t_c + t_j\,,
\end{align}
where $d_L$ and $t_c$ are the true luminosity distance and coalescence time.

%%%--------------------------------------------------------------------------
\section{The Strong lensing hypothesis}
\label{sec:StrongLensHypo}
Under the lensed hypothesis $\mathcal{H}_L$, when a GW from an image $j$ enters an interferometer, the data stream
\begin{equation}\label{eq:DataStream}
    d_j(t) = n_j(t) + h_{L}^j(t; \boldtheta, \boldLambda_j)\,
\end{equation}
where $n_j(t)$ is the detector noise contribution, and $h_{L}^j(t; \boldtheta, \boldLambda_j)$ is the lensed GW as seen by the detectors. 
For two observed lensed images, the two different data streams containing a lensed signal are connected through the binary parameters $\boldtheta$, such that the joint evidence (neglecting selection effects)~\citep{Liu:2020par, Lo:2021nae}
\begin{equation} \label{eq:lensed_evidence}
\begin{split}
    p(d_1, d_2| \mathcal{H}_L) = \int &p(d_1|\boldtheta, \boldLambda_1) p(d_2|\boldtheta, \boldLambda_2)  \\
    &\times p(\boldtheta, \boldLambda_1, \boldLambda_2) d\boldtheta d\boldLambda_1 d\boldLambda_2\,,
\end{split}
\end{equation}
where $p(d_1|\boldtheta, \boldLambda_1)$ and $p(d_2|\boldtheta, \boldLambda_2)$ are the individual likelihoods \citep{Veitch:2009hd} and $p(\boldtheta, \boldLambda_1, \boldLambda_2)$ is the prior. 

%%%--------------------------------------------------------------------------
Under the usual non-lensed hypothesis $\mathcal{H}_U$, the  data streams take the form
\begin{equation}\label{eq:DataStreamHU}
    d_j(t) = n_j(t) + h_{U}^j(t; \boldtheta_j)\,,
\end{equation}
where $\boldtheta_j$ are the binary parameters of the $j^{th}$ waveform, and $h_{U}^j(t; \boldtheta_j)$ is the waveform projected onto the detector frame. 
Under the unlensed hypothesis, the data streams contain unrelated signals, such that the joint evidence~\citep{Liu:2020par, Lo:2021nae} 
\begin{equation} \label{eq:unlensed_evidence}
    p(d_1, d_2| \mathcal{H}_U) = \int p(d_1|\boldtheta_1) p(d_2|\boldtheta_2) p(\boldtheta_1) p(\boldtheta_2) d\boldtheta_1 d\boldtheta_2\,,
\end{equation}
where $p(\boldtheta_{1})$ and $p(\boldtheta_{2})$ are the usual priors. 

Therefore, to test the lensed hypothesis, we adopt the ratio of evidences, or \emph{``coherence ratio''} 
\begin{align}\label{eq:2imageCoherence}
    \mathcal{C}^L_U &= \frac{p(d_1, d_2| \mathcal{H}_L)}{p(d_1, d_2| \mathcal{H}_U)} \,,
\end{align}
which describes how (dis)similar the two signals are. 
Note that here we do not call $\mathcal{C}^L_U$ a Bayes factor because it does not include selection effects~\citep{Lo:2021nae} and it is sensitive to the binary black hole population prior.\footnote{We refer the reader to~\citep{Lo:2021nae} for a detailed discussion. A brief note about selection effects and astrophysical priors can be found in Appendix~\ref{app:selection_note}. } 

%%%--------------------------------------------------------------------------
\section{The conditioned evidence}
\label{sec:ConditionedEvidence}

In this work, instead of evaluating the full joint evidence above, we evaluate the \emph{conditioned} evidence and the individual evidences to obtain the coherence ratio. This way we 
can accelerate the computation of the conditioned evidence using importance sampling and a lookup table, as will be explained below. 

%%%--------------------------------------------------------------------------
First, we can rewrite the joint evidence as (see Appendix~\ref{sec:DetailBayesianFramework})
\begin{equation}\label{eq:lensedLikeli}
    p(d_{1}, d_{2}|\mathcal{H}_{L}) = p(d_{1}|\mathcal{H}_L) p(d_{2}|d_{1}, \mathcal{H}_{L} ) \,.
\end{equation}
Defining the relative magnification between images $\mu_{\rm rel}$, the relative time delay between images $t_{21}$, and the relative Morse factor between images $n_{21}= n_2-n_1$,
the conditioned evidence 
\begin{equation}\label{eq:lensed_conditioned_evidence}
\begin{split}
    p(d_2|d_1, \mathcal{H}_L) = \int & \int p(d_2|\mathbf{\Theta}, \mathbf{\Phi}) p(\mathbf{\Theta}|d_1) d\mathbf{\Theta} \\ &\times p(\mathbf{\Phi}) d\mathbf{\Phi} \,,
\end{split}
\end{equation}
where $\mathbf{\Phi}=\{\mu_{\rm rel}, t_{21}, n_{21}\}$ are the relative image properties and $\mathbf{\Theta}$ are the effective parameters which absorb the lensing magnification into an observed luminosity distance and the lensing time delay into an observed coalescence time (see Eq.~(\ref{eq:effective_params})) and includes also the Morse factor of the first image $n_1$. 
That is, the usual prior in the evaluation of the second event has been replaced by the posterior of the first event $p(\mathbf{\Theta}|d_1)$ and the first event is fully described by $\mathbf{\Theta}$. 
The second event can be related to the first event through the difference in the image properties 
\begin{equation}
\begin{split}
    &t_c^{\rm obs, 2} = t_c^{\rm objs,1}+ t_{21}\,, \\
    &d_L^{\rm obs, 2} = \sqrt{\mu_{\rm rel}}\, d_L^{\rm objs,1} \,, \\
    &n_2 = n_1 + n_{21} \,.
\end{split}
\end{equation}
 
 %%%--------------------------------------------------------------------------
In terms of the conditioned evidence, the coherence ratio
\begin{align}
    \mathcal{C}^L_U &= \frac{p(d_1| \mathcal{H}_L)}{p(d_1| \mathcal{H}_U)} \frac{p(d_2| d_1, \mathcal{H}_L)}{p(d_2| \mathcal{H}_U)} \,.
\end{align}
Thus, instead of evaluating the full joint evidence, we can evaluate the conditioned evidence and the individual evidences to obtain the coherence ratio. 

The main desirable attribute of the conditioned evidence is that the integral converges faster than a regular joint parameter estimation run, as the posterior of the first event $p(\mathbf{\Theta}|d_1)$, which replaces the usual prior in the conditioned evidence, is already concentrated around the relevant, shared detector-frame parameters. 
Indeed, the integral has reduced partially into an importance sampling problem over the shared waveform parameters. 

%%%--------------------------------------------------------------------------
\section{Evaluating the conditioned evidence}
\label{sec:BayesianFrame}
There are, in principle, several ways to solve the conditioned evidence. 
Here we solve the conditioned evidence is by re-writing the conditioned evidence in terms of a "marginalised" likelihood
%%%--------------------------------------------------------------------------

\begin{equation}\label{eq:lensed_conditioned_evidence_approx}
    p(d_2|d_1, \mathcal{H}_L) = \int L(\mathbf{\Phi}) p(\mathbf{\Phi}) d\mathbf{\Phi}\,,
\end{equation}
where 
\begin{equation}
    L(\mathbf{\Phi}) =  \left\langle p(d_2|\mathbf{\Theta}, \mathbf{\Phi}) \right\rangle_{p(\mathbf{\Theta}|d_{1})}\,,
\end{equation}
the likelihood of the second event averaged over the posterior samples of the first event.\footnote{Choosing $d_1$ to be the event with a better-constrained posterior of the two events will ensure faster convergence of the marginalized likelihood.}
%%%--------------------------------------------------------------------------

Evaluating the conditioned evidence instead of the full joint evidence is desirable when it comes to speed, as the posterior $p(\mathbf{\Theta}|d_{1})$ is already concentrated around the relevant waveform parameters, which accelerates the convergence of the integral.
%%%--------------------------------------------------------------------------
Furthermore, by recycling the posterior samples of the first event\footnote{To be able to use the posteriors of the first event, the analysis of the first image has been adapted to incorporate Morse factor as a discrete parameter}, we can evaluate the marginalised likelihood without having to generate new trial GW waveforms, through a lookup table. 
The lookup table further accelerates the computation (see Appendix~\ref{sec:fast_Likelihood} for details). 

%%%----------------------------------------------------------------------

Once we have the conditioned evidence, we have access to the posterior distributions of the lensing parameters $p(\mathbf{\Phi} | d_{1}, d_{2})$. 
We can then reweigh the posterior to obtain (Appendix~\ref{sec:DetailBayesianFramework})
\begin{equation}\label{eq:JointReweight}
p(\mathbf{\Theta}, \mathbf{\Phi} | d_{1}, d_{2}) \propto \frac{p(d_{2} | \mathbf{\Theta}, \mathbf{\Phi})}{p(d_{1}, d_{2} | \mathbf{\Phi})}p(\mathbf{\Theta}|d_{1})p(\mathbf{\Phi} | d_{1}, d_{2}) \, .
\end{equation}
%%%%-------------------------------------------------------------------------------
This way, we can recover the joint parameters given by the combined observation; see Fig.~\ref{fig:ReweighedSkyLoc} for an illustration.

We note that Eqs.~\ref{eq:lensed_conditioned_evidence}~and~\ref{eq:lensed_conditioned_evidence_approx} are exact relationships. 
However, in evaluating them, we have limited the number of samples drawn from the posterior of the first event, to make the construction of the lookup table for each sample feasible memory-wise. 

%%%--------------------------------------------------------------------------
We have implemented our methodology as a module to the \textsc{Bilby} GW parameter estimation tool~\citep{Ashton:2018jfp}, 
and named it \textsc{Golum} (\emph{Gravitational-wave analysis Of Lensed and Unlensed waveform Models}). 
Since we apply the Morse factor to the overall waveform as in  Eq.~(\ref{eq:lensed_waveform}), we can use \textsc{IMRPhenomPv2}~\citep{Khan:2015jqa} (a precessing-spin waveform model), \textsc{IMRPhenomD}~\citep{Khan2016} (an aligned-spin waveform model) and other waveforms with precession or higher order modes in our analysis. 
In our trial analyses, a single evaluation of the conditioned evidence takes less than $\mathcal O(1)$ CPU hour\footnote{This is the total time to compute the evidence for the lensed image with a \emph{Intel(R) Core(TM) i7-9750H CPU @ 2.60GHz} processor.}, adopting the \textsc{PyMultinest} sampler~\citep{Feroz:2008xx, Feroz:2013hea}.

%%%-----------------------------------------------------------------------------------
\section{Example analysis}
\label{sec:TypicalAnalysis}

%%%----------------------------------------------------------------------------------
As a practical example, we inject the signal from a spinning, precessing binary black hole merger generated with \textsc{IMRPhenomPv2}, with parameters listed in the second column of Table~\ref{tab:InjParams}, into synthetic stationary, Gaussian noise for a network of 3 detectors (LIGO-Livingston, LIGO-Hanford, and Virgo) at design sensitivity \citep{aLIGOdesign, TheVirgo:2014hva}. 
This event has a network SNR of $\sim 23$. 
We then inject the event's lensed counterpart image, with relative magnification between the two events $\mu_{\mathrm{rel}} = 2$, relative Morse factor $n_{21} = 0.5$, and relative time delay $t_{21} = 14$ hr.
Throughout this analysis, we use a uniform prior for the relative magnification ($\mu_{\rm rel}\in [0.01,20]$), the time-delay ($t_{21} \in [t_{21}-0.1, t_{21}+0.1]\, \rm s$), the chirp mass ($\mathcal{M}_{c} \in [10, 100]\,M_{\odot}$ , with $\mathcal{M}_{c} = (m_{1}m_{2})^{3/5}/(m_{1}+m_{2})^{1/5}$ where $m_{1 ,2}$ are the component masses), and mass ratio ($q \in [0.1, 1]$, with $q = m_{2}/m_{1}$); the spin distribution is isotropic. The prior on the Morse factor is a discrete uniform distribution over the three possible values ($n_{1} \in \{0, 0.5, 1\}$) and the prior in Morse factor difference is a discrete uniform distribution over the four possible values ($n_{21} \in \{0, 0.5, 1, 1.5 \}$)\footnote{The negative value $n_{21} = -0.5$ is equivalent to the transformation $n_{21} = 1.5$, and the negative value $n_{21} = -1$ is equivalent to $n_{21} = 1$. Thus, we do not consider them.}. 
These priors do not include results from lens modelling of the inferred astrophysical population of binary black holes.

%%%---------------------------------------------------------------------------
\begin{table*}
\begin{minipage}{\textwidth}
    \centering
    \begin{tabular}{l l l}
        \hline
        \hline
        \textbf{Parameter} &  \textbf{Value lensed event} & \textbf{Value unlensed event} \\
        \hline
        \hline
         Mass 1 ($m_{1}$) & $36.0\,M_{\odot}$ & $35.8\,M_{\odot}$\\
         Mass 2 ($m_{2}$) & $29.2\,M_{\odot}$ & $11.4\,M_{\odot}$ \\
         Spin amplitude 1 ($a_{1}$) & 0.4 & 0.3 \\ 
         Spin amplitude 2 ($a_{2}$) & 0.3  & 0.2\\
         Tilt angle 1 ($tilt_{1}$) & 0.5 & 1.0 \\
         Tilt angle 2 ($tilt_{2}$) & 1.0 & 2.2 \\
         Spin vector azimutal angle ($\phi_{12}$) & 1.7 & 5.1  \\
         Precession angle about angular momentum ($\phi_{jl}$) & 0.3 & 2.5 \\
         Luminosity distance ($d_{L}$)\footnote{In the lensing framework, the distance of the event is the \emph{apparent} one, as both images are in fact affected by the lensing parameters.} & $1500$  Mpc & $500$ Mpc \\
         Inclination angle ($\iota$) & 0.4 & 1.9 \\
         Wave polarization ($\psi$) & 2.659 & 2.7 \\
         Unlensed phase of coalescence ($\phi_{c}$) & 1.3 & 3.7 \\
         Morse factor ($n_{1}$) & 0.5 & 0\\
         Right ascension ($\alpha$) & 1.375 & 3.9 \\
         Declination ($\delta$) & -1.2108 & 0.22 \\
         Time of coalescence ($t_{c}$) & 1126259642.413 & 10.04 \\
         \hline
         
    \end{tabular}
    \caption{Summary of the injection parameters used for the examples in Sec.~\ref{sec:TypicalAnalysis}. In this table and throughout the work, the angles are measured in radians.}
    \label{tab:InjParams}
    \end{minipage}
\end{table*}

\begin{figure}
    \centering
    \includegraphics[keepaspectratio, width=0.5\textwidth]{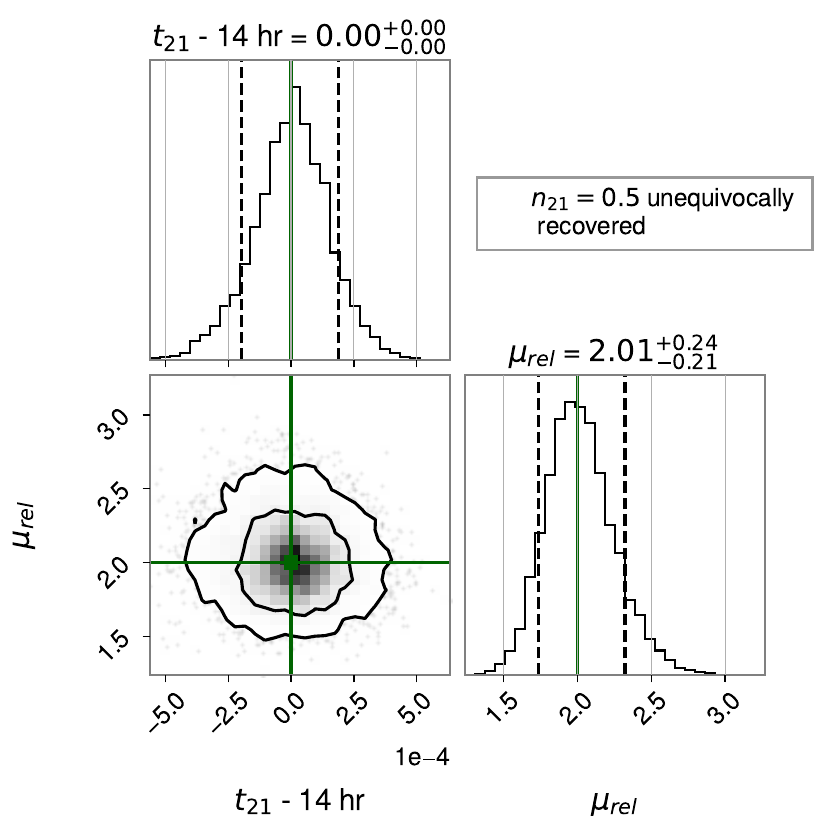}
    \caption{Posterior distribution of the magnification, and time delay (re-centered at zero in this figure) between two strongly lensed gravitational-wave images. The parameters are well recovered, and the difference in Morse factor is fully determined, allowing us to make inferences about the image properties. If the event pair were a part of a strongly lensed quadruplet lensed by a galaxy, the first image type would likely be Type-I and the second one Type-II. 
    The injected values are $\mu_{\mathrm{rel}} = 2$, $n_{21} = 0.5$, and $t_{21} = 50400\,\mbox{s} = 14\, \mbox{hr}$.}
    \label{fig:CornerLensing}
\end{figure}

%%%-----------------------------------------------------------------------------------
To analyse this lensed pair, we perform four nested sampling runs. 
Firstly, we analyse the two injections under the non-lensed hypothesis. 
Secondly, we estimate the parameters of one of the events under the lensing hypothesis. 
Thirdly, we obtain the conditioned evidence $p(d_2|d_1,\mathcal H_L)$, by sampling the second event's likelihood based on the earlier lensed parameter estimation, thereby also obtaining the relative image properties (see Fig.~\ref{fig:CornerLensing}).

%%%-----------------------------------------------------------------------------------
%%%--------------------------------------------------------------------------------------
Combining the four runs, we obtain the coherence ratio $\mathcal{C}^{L}_{U}$. 
In our example lensed simulation, we find $\log{C^{L}_{U}} = 23.6$, correctly consistent with lensing.
%%%-----------------------------------------------------------------------------------------
We then inject two events that are unrelated (see Table~\ref{tab:InjParams} for the parameters) and repeat the analysis. In this case, the coherence ratio $\log\mathcal{C}^{L}_{U} = -14 $, not consistent with lensing.

\begin{figure}
    \centering
    \includegraphics[keepaspectratio, width=0.5\textwidth]{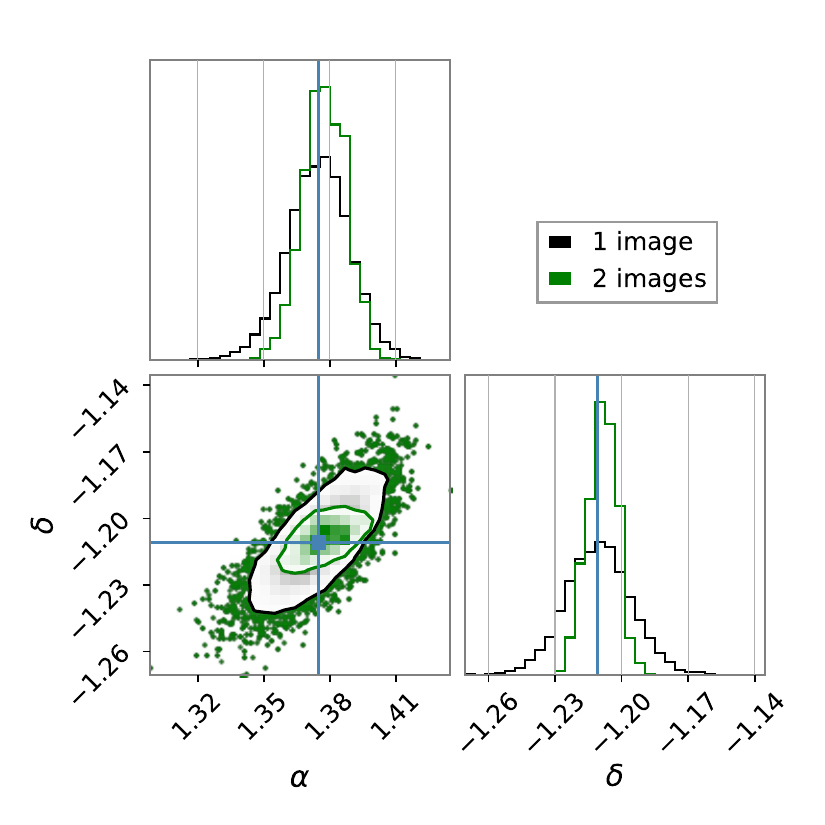}
    \caption{The combined sky location (right ascension $\alpha$ and declination $\delta$) of two strongly lensed gravitational-wave images. The black distributions refer to the posteriors from the analysis of the first image only, and the green distributions are the results when the two images are combined. The detection of several images significantly reduces the 90\% credible region.
    }
    \label{fig:ReweighedSkyLoc}
\end{figure}

%%%-----------------------------------------------------------------------------------------
We can also combine information from the two lensed images to better constrain the binary parameters.  
In particular, we can use the posterior of the lensed parameters that we obtain from the combined run to reweigh the posterior samples of the first run as in Eq.~(\ref{eq:JointReweight}). 
The most notable impact is on the sky localisation, where the 90\% confidence sky area improves by about a factor of two in our example case (Fig.~\ref{fig:ReweighedSkyLoc}). 
This is particularly important for the strong lensing science case, as an improved sky localisation might help narrow down the number of possible host galaxies when combining the GW information with electromagnetic data~\citep{Hannuksela:2020xor}. 

%%%--------------------------------------------------------------------------------------

%%%--------------------------------------------------------------------------
As an additional example, we analyse a sub-threshold trigger (a signal hidden in the noise background). 
In a targeted sub-threshold search, one uses a template bank to cover the source parameters posteriors recovered from the primary super-threshold event~\citep{Li:2019osa, McIsaac:2019use} -- such searches may uncover many additional candidates, which would need to be analysed. 
We assume that a super-threshold event has already been observed (with identical parameters to the event described in the earlier example). 
The lensing parameters for this event are $\mu_{\mathrm{rel}} = 25$\footnote{For this search, the prior for the relative magnification was extended to cover the $[0.01, 50]$ interval.}, $n_{2} = 1$, and $t_{21} = 16\, \rm hr$, leading to a network SNR of $\sim 5.5$, which is below the value typically required for detection. 

\begin{figure}
    \centering
    \includegraphics[keepaspectratio, width=0.5\textwidth]{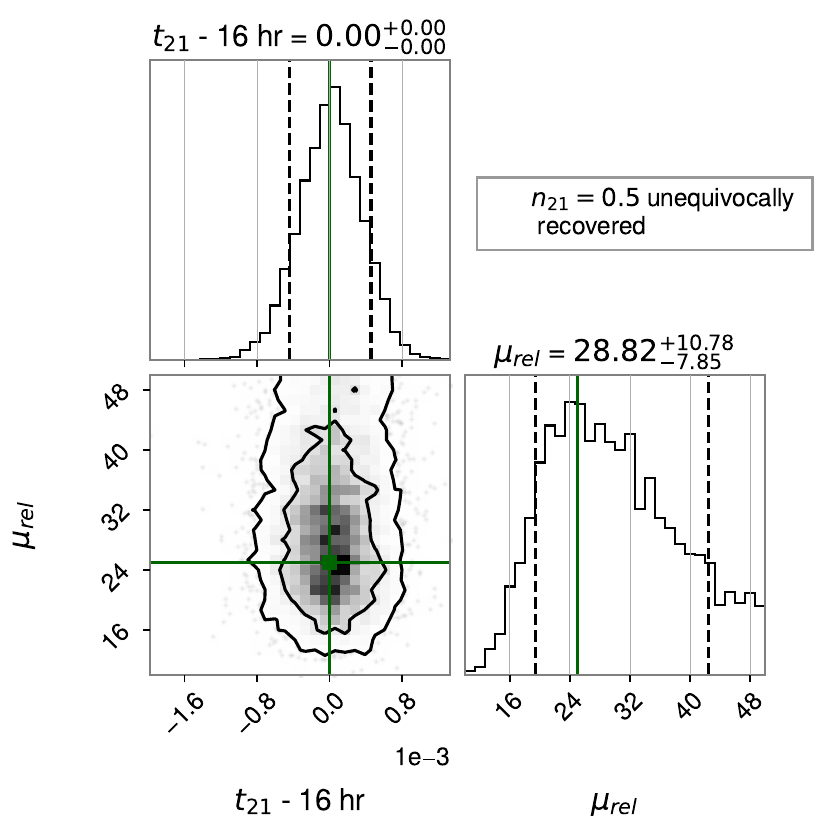}
    \caption{
    Posterior distribution of the  magnification, and time delay (re-centered at zero) between two strongly lensed gravitational-wave images, where one of the events is sub-threshold (with an SNR of $\sim 5.5$). 
    The simulated relative magnification $\mu_{\mathrm{rel}} = 25$, relative Morse factor $n_{21} = 0.5$, and relative time delay $t_{21}= 57600 \, \text{s} = 16\, \text{hr}$. 
    The parameters are well recovered and the Morse factor difference is also uniquely recovered in this scenario.}
    \label{fig:SubThreshCorner}
\end{figure}

%%%--------------------------------------------------------------------------
The resulting posteriors are shown in Fig.~\ref{fig:SubThreshCorner}. We recover injected values, but (as expected) the relative magnification measurement is less accurate than it is for typical super-threshold events. The coherence ratio $\log\mathcal{C}^{L}_{U} = 9.3$, consistent with lensing. 

%%%---------------------------------------------------------------------------------------------------

%%%--------------------------------------------------------------------------
\section{Multiple image analyses}
\label{sec:MultipleImages}
In practice, strong lensing will often produce more than two images~\citep{1992SchneiderBook}. 
Our approach enables one to study multiple images within a Bayesian framework in a computationally tractable way. The methodology detailed below is applicable to any number of images produced by strong lensing (see also~\citep{Lo:2021nae} for a derivation in the context of full joint parameter estimation). 

%%%--------------------------------------------------------------------------
Under the hypothesis that $N$ individual GW events are lensed, their detector-frame parameters are related to one another, while, under the unlensed hypothesis, they are not. 
Accounting for multiple images, the coherence ratio 
\begin{align}\label{eq:MultiImageCoheranceRatio}
    \mathcal{C}^{L}_{U} &= \frac{p(d_{1}, \dots, d_{N} | \mathcal{H}_{L}) }{p(d_{1}, \dots, d_{N} | \mathcal{H}_{U})}\,, 
\end{align}
with the numerator 
\begin{eqnarray}\label{eq:ConditionedEvidencNimgs}
    &&p(d_{1}, \dots, d_{N} | \mathcal{H}_{L}) \nonumber\\ 
    &=& p(d_{1} | \mathcal{H}_{L}) \prod_{i = 2}^{N} p(d_{i} | d_{1}, \dots, d_{i-1},  \mathcal{H}_{L}) \,,
\end{eqnarray}
and the denominator
\begin{equation}\label{eq:UnlensedHypoNimg}
    p(d_{1}, \dots, d_{N} | \mathcal{H}_{U}) = \prod_{i = 1}^{N} p(d_{i} | \mathcal{H}_{U})\,.
\end{equation}

%%%--------------------------------------------------------------------------
The conditioned evidence for the $i^{th}$ image can then be solved similarly to the case with two images (see Eq.(\ref{eq:lensed_conditioned_evidence_approx})):
\begin{equation}\label{eq:LensedCondi_Nimages}
    p(d_{i} | d_{1}, \dots, d_{i-1}, \mathcal{H}_{L}) = \int L_{i}(\mathbf{\Phi}_i)p(\mathbf{\Phi}_i) d\mathbf{\Phi}_i \,,
\end{equation}
where $\mathbf{\Phi}_i$ represents the lensing parameters for the $i^{th}$ image, i.e the parameters linking the first observed image to the one under consideration. In this expression, the marginalized likelihood is obtained by taking the average over the reweighed samples resulting from the run for the $(i-1)^{th}$ image: 
\begin{eqnarray}\label{eq:GeneralizedMarginalLikelihood}
    &&L_{i}(\mathbf{\Phi}_i) \nonumber\\
    &=& \langle p(d_{i} | \mathbf{\Theta}, \mathbf{\Phi}_i, n_{2}, \dots, n_{i-1}) \rangle_{p(\mathbf{\Theta}, n_{2}, \dots, n_{i-1} | d_{1}, \dots, d_{i-1})}. 
    \nonumber\\
\end{eqnarray}
Finally, the combined posterior distribution
\begin{equation}\label{eq:generalizedResamp}
\begin{split}
    p(\mathbf{\Theta}, \mathbf{\Phi}_i, n_{2}, \dots n_{i-1} & | d_{1}, \dots, d_{i}) \\  & \propto \frac{p(d_{i} | \mathbf{\Theta}, \mathbf{\Phi}_i, n_{2}, \dots n_{i-1})}{L_{i}(\mathbf{\Phi}_i)} \\ & \times p(\mathbf{\Theta}, n_{2}, \dots n_{i-1} | d_{1}, \dots d_{i-1})  \\ & \times p(\mathbf{\Phi}_i | d_{1}, \dots, d_{i}).
\end{split}
\end{equation}
Indeed, to analyse a set of images, one uses the posterior of the first image as a prior for the analysis of the second image; then, the posterior of the combined first two events for the analysis of the third image; \emph{ad infinitum} until all images have been analysed. 

%%%--------------------------------------------------------------------------
\section{Quadruple image analysis: Sky localisation}
\label{sec:FourImageExample}
Let us analyse an example quadruplet of lensed images. 
We assume that the first and second images have the same parameters as in Sec.~\ref{sec:TypicalAnalysis}. 
We inject two more lensed signals, with relative magnifications of $4$ and $5$, time delays of $16$ hours and $21$ hours, and Morse factors $n_{3} = 0$ and $n_{4} = 1$, respectively.

%%%--------------------------------------------------------------------------
\begin{figure}
    \centering
    \includegraphics[keepaspectratio, width=0.5\textwidth]{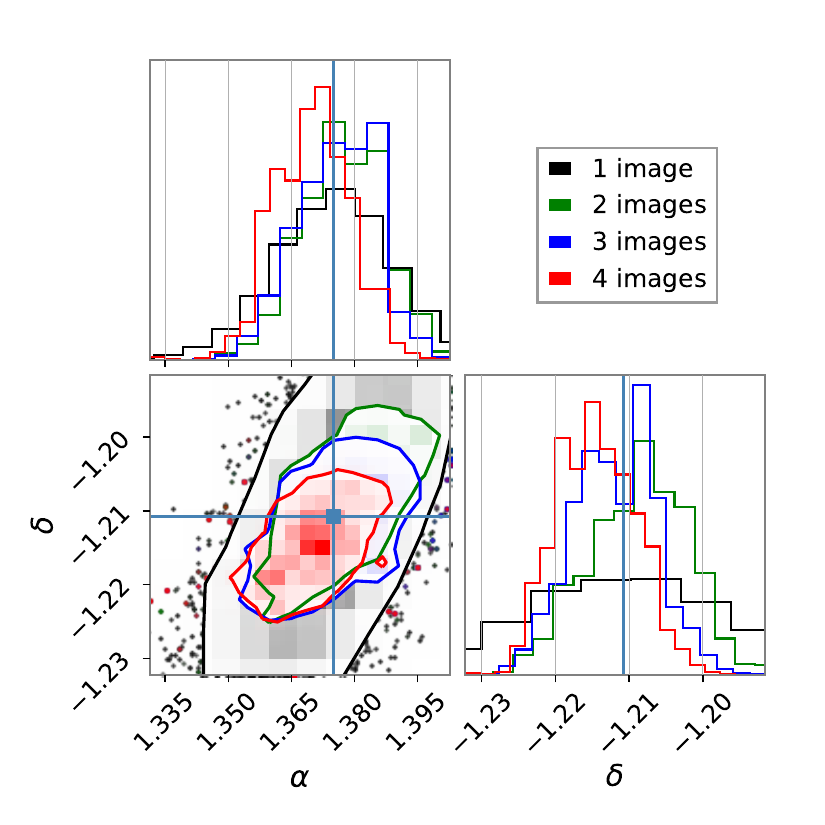}
    \caption{The 90\% credible region of the sky location of a strongly lensed gravitational-wave singlet (black), pair (green), triplet (blue),  and quadruplet (red). 
    There is a clear improvement in the sky localization with the addition of every gravitational-wave image. 
    The final 90\% confidence sky area is $\sim 2\deg^{2}$ in this example.
    An improved sky localization might be particularly useful for lensed host galaxy localization. 
    }
    \label{fig:4ImgSkyLoc}
\end{figure}

%%%--------------------------------------------------------------------------
We begin by analysing the first two images. 
We then use the reweighed samples obtained from the joint analysis of the first two images to analyse the third image. 
As a consequence, we retrieve the lensing parameters $p(\mathbf{\Phi}_3|d_1,d_2,d_3,\mathcal{H}_L)$ and the conditioned evidence $p(d_3|d_1,d_2,\mathcal{H}_L)$ for the third image. 
We then reweigh the posterior samples from the second run with the results from third, obtaining $p(\mathbf{\Theta},\mathbf{\Phi}_3|d_1,d_2,d_3)$. 
Using those reweighed samples, we analyse the fourth image similarly, obtaining $p(\mathbf{\Theta},\mathbf{\Phi}_4|d_1,d_2,d_3,d_4)$. 

%%%--------------------------------------------------------------------------
A particularly noteworthy improvement is in the sky localization, which we show in Fig.~\ref{fig:4ImgSkyLoc}. 
The initial $90\%$ sky area of $\sim 20\,\rm deg^{2}$ of the first image is reduced to a final area of $\sim 2 \, \rm deg^{2}$ when accounting for the four images. 
Such an improvement is particularly important for studies involving lensed host galaxy localisation, which rely on an accurate sky map estimate~\citep{Hannuksela:2020xor}.
The entire analysis was performed in around $12$ CPU hours\footnote{This corresponds to the time needed to infer the lensed parameters and reweigh the posteriors for the three images}. 

%%%----------------------------------------------------------------------------------
\section{Conclusions}
\label{sec:Conclusions}
This work introduces an approximate joint parameter estimation methodology, which allows for fast and precise strong lensing analyses. 
We have demonstrated its use in the analysis of simulated lensed and unlensed events. 
 
Such a methodology enables us to analyse lensed events at a relatively high accuracy with low computational cost. 
Furthermore, it enables quick multiple-image analyses. 
The combination of speed and precision allowed by our method will likely become crucial in the future when we expect the number of detected individual events, each of which could, in principle, be a lensed image, to rise rapidly. 
Future work may focus on more realistic population priors and selection effects~\citep{Lo:2021nae}. 
We anticipate this methodology to play a complementary role to the existing posterior-overlap~\citep{Haris:2018vmn} and joint parameter estimation~\citep{Liu:2020par, Lo:2021nae} methodologies, where its role would be to perform strong lensing estimates and multiple-image analyses in an accelerated fashion -- situated between the two existing methodologies in terms of speed and precision. 
A three-tier analysis may be possible in the future, where we first analyse the strongly lensed events with a posterior-overlap method, after which we analyse a reduced set of events with our methodology, and finally, the best candidate(s) could be passed to the joint parameter estimation. 
%%%---------------------------------------------------------------------------------------

\section*{Acknowledgements}
\label{sec:Acknow}
The authors thank  Rico Ka Lok Lo, and Chinmay Kalaghatgi for their useful comments and feedback throughout this work. 
We also thank Tjonnie Li, Eungwang Seo, Isaac Wong, Leon Koopmans, Ewoud Wempe, and Renske Wierda for useful discussion and insight; we thank Thomas E. Collett for discussion on related projects. 
J.J, O.A.H, K.H, and C.V.D.B. are supported by the research program of the Netherlands Organisation for Scientific Research (NWO).
The authors are grateful for computational resources provided by the LIGO Laboratory and supported by the National Science Foundation Grants No. PHY-0757058 and No. PHY-0823459.  

\section*{Data availability}
The data underlying this article will be shared in reasonable request to the corresponding authors.

\bibliography{bibly}
%%%--------------------------------------------------------------------------
\onecolumn

\appendix
%%%--------------------------------------------------------------------------
\section{Details of the Bayesian framework}
\label{sec:DetailBayesianFramework}

\subsection{Marginalized likelihood}
Denoting $\mathbf{\Theta}$ as the set of biased parameters and Morse factor for the reference image, and $\mathbf{\Phi}$ as the set of lensed parameters, the joint likelihood under the strong lensing hypothesis
\begin{align}\label{eq:AppendixlensedLikeli}
    p(d_{1}, d_{2}|\mathcal{H}_{L}) &= \int \int p(d_1|\mathbf{\Theta}, \mathcal{H}_{L})p(d_{2}|\mathbf{\Theta}, \mathbf{\Phi}, \mathcal{H}_{L}) p(\mathbf{\Theta} | \mathcal{H}_{L}) p(\mathbf{\Phi}|\mathcal{H}_{L}) d\mathbf{\Theta} d\mathbf{\Phi}  \nonumber \\
    &= p(d_{1}|\mathcal{H}_{L}) \int \int p(\mathbf{\Theta}|d_{1}, \mathcal{H}_{L})p(d_{2}|\mathbf{\Theta}, \mathbf{\Phi}, \mathcal{H}_{L})P(\mathbf{\Phi}|\mathcal{H}_{L})d\mathbf{\Theta} d\mathbf{\Phi} \nonumber \\
    &= p(d_{1}|\mathcal{H}_{L}) \int \int p(d_{2}, \mathbf{\Theta}, \mathbf{\Phi}| d_{1},\mathcal{H}_{L})d\mathbf{\Theta} d\mathbf{\Phi} \nonumber \\
    &= p(d_{1}|\mathcal{H}_{L}) p(d_{2}| d_{1}, \mathcal{H}_{L})\,.
\end{align}
The likelihood under the unlensed hypothesis 
\begin{equation}\label{eq:AppendincUnlensedLikeli}
    p(d_{1}, d_{2} | \mathcal{H}_{U}) = p(d_{1} | \mathcal{H}_{U})p(d_{2} | \mathcal{H}_{U}).
\end{equation}
Inserting the joint likelihood expressions into the expression for the coherence ratio, we obtain 
\begin{align}\label{eq:AppendixBayesFactorSimp}
    \mathcal{C}^{L}_{U} &= \frac{p(d_{1}| \mathcal{H}_{L})}{p(d_{1} | \mathcal{H}_{U})}\frac{p(d_{2}|d_{1}, \mathcal{H}_{L})}{p(d_{2}|\mathcal{H}_{U})}
\end{align}
The conditioned evidence 
\begin{equation}\label{eq:AppendixApproxProba}
\begin{split}
    p(d_{2}|d_{1}, \mathcal{H}_{L}) &= \int \int p(d_{2}, \mathbf{\Theta}, \mathbf{\Phi} |d_{1}, \mathcal{H}_{L}) d\mathbf{\Theta} d\mathbf{\Phi}  \\
    &= \int p(\mathbf{\Phi}|\mathcal{H}_{L})  \int p(d_{2}|\mathbf{\Theta}, \mathbf{\Phi}, \mathcal{H}_{L}) p(\mathbf{\Theta}|d_{1} , \mathcal{H}_{L}) d\mathbf{\Theta} d\mathbf{\Phi}   \\
    &= \int \left\langle L_{2}(\mathbf{\Theta}, \mathbf{\Phi}) \right\rangle_{p(\mathbf{\Theta}|d_{1}, \mathcal{H}_{L})} p(\mathbf{\Phi}|\mathcal{H}_{L}) d\mathbf{\Phi}  \\
    &= \int L(\mathbf{\Phi}) p(\mathbf{\Phi}|\mathcal{H}_{L}) d\mathbf{\Phi}\,.
\end{split}
\end{equation}
This approximation reduces the 19D (assuming generic spins) integral to a 3D integral by using the posterior distributions of the first event for the source parameters. This expression is the key to our importance sampling method, giving rise to a significant speed-up.

%%%--------------------------------------------------------------------------
\subsection{Development for the posterior re-sampling}
The probability to have both the source and lensing parameters
\begin{equation}
p(\mathbf{\Theta}, \mathbf{\Phi} | d_{1}, d_{2}) = p(\mathbf{\Theta} | \mathbf{\Phi}, d_{1}, d_{2})p(\mathbf{\Phi} | d_{1}, d_{2}) \, .
\end{equation}

%%%--------------------------------------------------------------------------
However, the probability to have the source parameters given the lensed parameters and the two data streams
\begin{align}
    p(\mathbf{\Theta} | \mathbf{\Phi}, d_{1}, d_{2}) &= \frac{p(d_{1}, d_{2} | \mathbf{\Theta}, \mathbf{\Phi})\,p(\mathbf{\Theta}|\mathbf{\Phi})}{p(d_{1}, d_{2} | \mathbf{\Phi})} \\
    &= \frac{p(d_{2} | d_{1}, \mathbf{\theta}, \mathbf{\Phi}, n_{1}) \, p(d_{1} | \mathbf{\Theta}, \mathbf{\Phi}) \, p(\mathbf{\Theta} | \mathbf{\Phi})}{p(d_{1}, d_{2} | \mathbf{\Phi})} \\
    & \propto \frac{p(d_{2} | \mathbf{\Theta}, \mathbf{\Phi}) p(\mathbf{\Theta} | d_{1})}{L(\mathbf{\Phi})} \, .
\end{align}
To go from the second to the third line, we used $p(d_{1}, d_{2} | \mathbf{\Phi}) = L_{marg}(\mathbf{\Phi})$ (which is the output of \textsc{GOLUM}), $p(d_{2} | d_{1}, \mathbf{\Theta}, \mathbf{\Phi}) = p(d_{2} | \mathbf{\Theta}, \mathbf{\Phi})$ (conditional independence of the second data set on the first given all the parameters), $p(d_{1} | \mathbf{\Theta}, \mathbf{\Phi} ) = p(d_{1} | \mathbf{\Theta}) $ (independence of the first data-steam on the lensing parameters), and $p(\mathbf{\Theta} | \mathbf{\Phi}) = p(\mathbf{\Theta})$ (independence of the source parameters on the lensing parameters).

%%%--------------------------------------------------------------------------
Inserting the second expression in the first,
\begin{equation}
p(\mathbf{\Theta}, \mathbf{\Phi} | d_{1}, d_{2}) \propto \frac{p(d_{2} | \mathbf{\Theta}, \mathbf{\Phi})}{L(\mathbf{\Phi})}p(\mathbf{\Theta}|d_{1})p(\mathbf{\Phi} | d_{1}, d_{2}) \, .
\end{equation}

%%%--------------------------------------------------------------------------
\section{Fast Likelihood Computation}
\label{sec:fast_Likelihood}
As described in Appendix~\ref{sec:DetailBayesianFramework}, $L(\mathbf{\Phi})$ is obtained by averaging $L_{2}(\mathbf{\Theta}, \mathbf{\Phi})$ over a subset of posteriors samples for $\mathbf{\Theta}$ taken from $p(\mathbf{\Theta}|d_{1})$.  The computation of  $L(\mathbf{\Phi})$  can be speed-up by decomposing the $L_{2}(\mathbf{\Theta}, \mathbf{\Phi})$ dependencies on $\mathbf{\Theta}$ and $\mathbf{\Phi}$. 

%%%--------------------------------------------------------------------------
Let $\left\{d^j_1\right\}$ and $\left\{d^j_2\right\}$ be the data sets corresponding to image 1 and image 2 respectively, with $j$ denoting the detector index. The log likelihood for the second image can be written as,
\begin{eqnarray}
\label{eq:ll1_1}
2 \ln L_2(\mathbf{\Theta},\mathbf{\Phi}) &=& \sum_j \langle d^j_2-h^j(\mathbf{\Theta},\mathbf{\Phi}) |d^j_2-h^{j}(\mathbf{\Theta},\mathbf{\Phi})  \rangle \nonumber \\
&=& \sum_j  \langle d^j_2|d^j_2 \rangle +  \frac{1}{\mu_{\mathrm{rel}}}\langle h^j(\mathbf{\Theta})|h^j(\mathbf{\Theta}) \rangle - 2 \frac{1}{\sqrt{\mu_{\mathrm{rel}}}}\ \langle d^j_2|h^j(\mathbf{\Theta}) e^{2\pi i f t_{21} - i\pi n_{21} \text{sign}(f)} \rangle\,,
\end{eqnarray}
where,
\begin{equation}
     \langle d^j_2|h^j(\mathbf{\Theta}) e^{2\pi i f t_{21} - i\pi n_{21} \text{sign}(f)} \rangle = \Re\left[  W_j(\mathbf{\Theta},t_{21} ) e^{i\pi n_{21} }\right]\,,
\end{equation}
with the complex SNRs $W_j$, 
\begin{equation}
\label{eq:cplx_snrs}
 W_j(\mathbf{\Theta}, t_{21}) \equiv   4 \int_{-\infty}^{\infty} \frac{\tilde d^j_2(f) \tilde h^{*}(\mathbf{\Theta},f)}{S^j{f}}  e^{-2\pi i f t_{21} } df  =  \mathscr{F}^{-1} \left[ \frac{\tilde d^j_1(f) \tilde h^{*}(\mathbf{\Theta},f)}{S^j(f)} \right] \,.
\end{equation}
The complex SNR time series $W_j$ and template scalar products $\langle h^j(\mathbf{\Theta})|h^j(\mathbf{\Theta}) \rangle$ can be pre-computed for the $\mathbf{\Theta}$ samples and saved as a lookup table. The likelihood of the second image, $L_2(\mathbf{\Theta},\mathbf{\Phi})$ can be computed by substituting entries from the lookup table into the Eq.~(\ref{eq:ll1_1}).

This procedure enables us to gain a considerable amount of time when analysing the second image, dividing the run-time by $\sim 20$.

%%%--------------------------------------------------------------------------------------------------------

\section{Note on selection effects and astrophysical priors} 
\label{app:selection_note}
In this work, we do not call $\mathcal{C}^L_U$ the Bayes factor because it does not include selection effects~\citep{Lo:2021nae}. It is also sensitive to the intrinsic population properties of binary black holes (mass and spin population), which are not necessarily sufficiently well-resolved to guarantee a robust interpretation of the Bayes factor (see, e.g., ~\citep{Dai:2020tpj}, for a discussion). 
Such selection effects were described in details for the first time in~\citep{Lo:2021nae} (full treatment of the binary black hole population priors is also given in the reference), and here we briefly summarise some of the results. 
For more details, we refer the reader to~\citep{Lo:2021nae}, which introduces \textsc{hanabi}, a joint parameter-estimation tool including selection effects and population prior modeling. 

%%%--------------------------------------------------------------------------
The Bayes factor, including selection effects~\citep{Lo:2021nae}
\begin{equation}
\begin{split}
    &p(d_1, d_2| \dett, \mathcal H_L) = \frac{p(d_1,d_2|\mathcal H_L)P(\dett | d_1, d_2, \mathcal{H}_{L})}{P(\dett|\mathcal H_L)} \,, \\
    &p(d_1, d_2| \dett, \mathcal H_U) = \frac{p(d_1,d_2|\mathcal H_U)P(\dett | d_1, d_2, \mathcal{H}_{U})}{P(\dett|\mathcal H_U)} \,, \\
    &\mathcal{B}^L_U = \mathcal{C}^L_U   \frac{P(\dett|d_1, d_2, \mathcal H_L) P(\dett|\mathcal H_U)}{P(\dett | d_1, d_2, \mathcal{H}_{U})P(\dett|\mathcal H_L)} = \mathcal{C}^L_U   \frac{P(\dett|d_1, d_2, \mathcal H_L) P(\dett|\mathcal H_U)}{P(\dett|\mathcal H_L)} \,,
\end{split}
\end{equation}
where $\dett$ conditions the likelihood with a detection. Note that in the unlensed hypothesis, one has $P(\dett | d_1, d_2, \mathcal{H}_{U}) =1$, since a detection has been made~\citep{Lo:2021nae, 2019MNRAS.486.1086M}.
Selection effects and systematic prior considerations requiring extensive injection studies are outside our study's scope. 
However, there are plans to include them through interfacing our package with \textsc{hanabi}, which accounts for these effects, in the future. 
\end{document}